\documentclass{amsart}
\usepackage{amsfonts}

\usepackage{graphicx}

\input{tcilatex}
\input tcilatex

\begin{document}
\title[On the spread of a branching Brownian motion]{On the spread of a branching Brownian motion whose offspring number has
infinite variance}
\author{Jean Avan, Nicolas Grosjean and Thierry Huillet}
\address{CNRS, UMR-8089 and University of Cergy-Pontoise\\
2, rue Adolphe Chauvin F-95302, Cergy-Pontoise, Cedex, FRANCE\\
E-mail(s): avan@u-cergy.fr, nicolas.grosjean@u-cergy.fr, huillet@u-cergy.fr}
\maketitle

\begin{abstract}
\textbf{We study the impact on shape parameters of
an underlying Bienaym\'{e}-Galton-Watson branching process (height,
width and first hitting time), of having a non-spatial branching mechanism with infinite variance.  
Aiming at providing a comparative study of the spread of an epidemics whose
dynamics is given by the modulus of a branching Brownian motion (BBM) we then consider 
spatial branching processes in dimension $d$, not necessarily integer.
The underlying branching mechanism
is then either a binary branching model or one presenting infinite
variance. In particular we evaluate the chance $p(x)$ of
being hit if the epidemics started away at distance $x.$ We compute the
large $x$ tail probabilities of this event, both when 
the branching mechanism is regular and when it exhibits 
very large fluctuations. }
\end{abstract}


\section{Introduction}

The aim of this paper is a comparative study of the spread of an epidemics
whose dynamics is given by the modulus of a branching Brownian motion (BBM)
in dimension $d,$ not necessarily integer; the underlying branching
mechanism is either the one of a binary branching model or the one
presenting infinite variance which we define presently.

Before the spatial aspects of the BBM are addressed, we study the impact of
having a branching mechanism with infinite variance on the shape of the
underlying continuous-time Bienaym\'{e}-Galton-Watson tree process. This
chiefly concerns the time to extinction (the height of the tree), the
maximum population size (the width of the tree) and its first hitting time.
We compute the laws of these shape quantities for both the binary and the
infinite variance branching mechanisms, in the sub-, super- and critical
regimes, and we compare the two situations. The obtained results are
developed in Section $2$.

In Section $3$, space is introduced. The special spatial BBM in dimension 
$d=1$ is addressed specifically since the model is then exactly
solvable. Following the work of \cite{Sawyer}, we study the probability $%
p\left( x\right) $ that the Eve particle starting at $0$ has some descendant
ever diffusing above the threshold $x\geq 0.$ The main new aspects of our
results concerns the branching mechanism with infinite variance and its
comparison with the binary branching model typically studied in \cite{Sawyer}
where the new individuals that come to birth along the branching mechanism
are viewed as new mutants in an infinite allele model of population genetics.

In Section $4$, we deal with the $d\neq 1$ case. We study $p(x)$, the probability that the Eve particle starting at a distance $x$ of the origin has any of its descendants ever diffusing within a ball of radius $\epsilon $ centered at the origin. It is found that $p(x)$ satisfies a non-linear differential equation, which we use to compute its tail probabilities. In the critical case, the equation exhibits exact conformal covariance and has a corresponding
invariant power-law solution.
The large $x$ behavior of $p(x)$ is then in any case power-law, the exponent of which depends sharply on the value of the dimension $d$ with respect to a critical dimension $d_{c},$ reflecting the very large fluctuations of the branching mechanism. In the sub- and super-critical case, the large $x$ behavior of $p(x)$ is exponential with a multiplicative power prefactor depending on dimension $d$.

\section{Branching processes: a reminder}

\subsection{Generalities and well-known facts}

Let us start with well-known facts on continuous-time elementary branching
Bienaym\'{e}-Galton-Watson (BGW) processes, \cite{Harris}.

Suppose at some random (mean one) exponential time, one initial individual
dies out and produces a random number $M$ of offspring, with $M\in \left\{
0,1,2,...\right\} $. Let $f\left( z\right) =\mathbf{E}\left( z^{M}\right) ,$ 
$z\in \left[ 0,1\right] $, be the probability generating function (pgf) of $%
M, f\left( z\right) =\sum_{k\geq 0}\pi _{k}z^{k}$ with $\pi _{k}=\mathbf{P}%
\left( M=k\right) $.

Let $\phi _{t}\left( z\right) =\mathbf{E}\left( z^{N_{t}}\right) $, $\phi
_{0}\left( z\right) =z,$ be the pgf of the number of particles $N_{t}$ alive
at time $t\geq 0$. Then, setting $g\left( z\right) =f\left( z\right) -z,$ $%
\phi _{t}\left( z\right) $ solves 
\begin{equation*}
\overset{\cdot }{\phi }_{t}\left( z\right) =g\left( \phi _{t}\left( z\right)
\right) ,\text{ }\phi _{0}\left( z\right) =z,
\end{equation*}
where the `$\overset{}{\overset{.}{}}$' represents partial differentiation
with respect to time.

We assume in the sequel that $\mu :=f^{^{\prime }}\left( 1\right) =\mathbf{E}%
\left( M\right) <\infty .$

If $\mu :=f^{^{\prime }}\left( 1\right) =\mathbf{E}\left( M\right) <1$, the
process is denoted ``subcritical''. It is supercritical if $\mu >1$ and critical if $\mu
=1$.

When the process is either critical or subcritical, extinction occurs with
probability $1$, meaning $N_{\infty }=0$; otherwise if it is supercritical,
extinction occurs with probability $\rho <1$ which is the smallest solution
to $f\left( \rho \right) =\rho $ ($g\left( \rho \right) =0$)$.$ Note that $%
f^{\prime }\left( \rho \right) <1$.  A supercritical process explodes ($N_{\infty
}=\infty $) with complementary probability $%
\overline{\rho }=1-\rho >0$

The probability that the time to extinction, say $\tau _{e},$ is smaller
than $t$ is $\phi _{t}:=\phi _{t}\left( 0\right) =\mathbf{P}\left(
N_{t}=0\right) =\mathbf{P}\left( \tau _{e}\leq t\right) $, solution to 
\begin{equation}
\overset{\cdot }{\phi }_{t}=g\left( \phi _{t}\right) ,\text{ }\phi _{0}=0.
\label{E2}
\end{equation}
Alternatively, the probability that the time to extinction $\tau _{e}$ is
larger than $t$ is $\overline{\phi }_{t}:=1-\phi _{t}\left( 0\right) =%
\mathbf{P}\left( N_{t}>0\right) =\mathbf{P}\left( \tau _{e}>t\right) $,
solution to 
\begin{equation*}
\overset{\cdot }{\overline{\phi }}_{t}=-h\left( \overline{\phi }_{t}\right) ,%
\text{ }\overline{\phi }_{0}=1,
\end{equation*}
where $h\left( z\right) =g\left( 1-z\right) $. $\tau _{e}$ is also called
the height of the BGW tree.

Whenever $M$ has all its moments finite, it holds that 
\begin{equation*}
h\left( z\right) =\left( 1-\mu \right) z+\sum_{k\geq 2}\mathbf{E}\left[
\left( M\right) _{k}\right] \frac{\left( -z\right) ^{k}}{k!},
\end{equation*}
where $\mathbf{E}\left[ \left( M\right) _{k}\right] :=\mathbf{E}\left[
M\left( M-1\right) ...\left( M-k+1\right) \right] $ are the falling
factorial moments of $M.$\newline

We now specialize to two main special BGW cases:\newline

\textbf{1.} (binary branching) $f\left( z\right) =\pi _{0}+\pi _{1}z+\pi
_{2}z^{2}.$ Here $h\left( z\right) =\left( 1-\mu \right) z+2\pi _{2}\frac{%
z^{2}}{2!},$ with $\mu =1-\left( \pi _{0}-\pi _{2}\right) $ and $\mu >1$ iff 
$\pi _{2}>\pi _{0}.$

\textbf{2.} (infinite variance Lamperti branching model \cite{Lamperti}): $%
f\left( z\right) =1-\mu \left( 1-z\right) +C\left( 1-z\right) ^{\gamma }$
where $\mu /\gamma >C>\mu -1$ and $\gamma \in \left( 1,2\right) $ so that $%
f\left( 0\right) \in \left( 0,1\right) .$ Here $h\left( z\right) =\left(
1-\mu \right) z+Cz^{\gamma }$ and since $\gamma \in \left( 1,2\right) $, the
variance of $M$ is infinite, in contrast with the preceding binary splitting
model. Given our constraints on $C$, $f\left( z\right) $ is a well-defined
completely monotone pgf (in particular, $f^{\prime }\left( z\right) >0$, for
all $z\in \left( 0,1\right) $).

The probability system for this model is 
\begin{equation}
\pi _{0}=1-\mu +C,\text{ }\pi _{1}=\mu -C\gamma ,\text{ }\pi _{k}=C\frac{%
\gamma \left( \gamma -1\right) \left( 2-\gamma \right) ...\left( k-\gamma
-1\right) }{k!}\text{, }k\geq 2.  \label{E1}
\end{equation}
\newline

Let us investigate $\overline{\phi }_{t}=\mathbf{P}%
\left( \tau _{e}>t\right) $ for these $2$ models. We compute the exact form of the
function and its asymptotic behavior at infinite time:\newline

Consider first \emph{Model} \textbf{1}.
It is easy to establish:

\begin{proposition}
-a) In the subcritical case $\mu <1$ and $\overline{\rho }=0:$

$\mathbf{P}\left( \tau _{e}>t\right) =e^{-\left( 1-\mu \right) t}/\left(
1+\pi _{2}\left( 1-e^{-\left( 1-\mu \right) t}\right) /\left( 1-\mu \right)
\right) \sim \left( 1+\pi _{2}/\left( 1-\mu \right) \right) ^{-1}e^{-\left(
1-\mu \right) t}$ with exponential tails.

-b) In the critical case $\mu =1$ and $\overline{\rho }=0:$

$\mathbf{P}\left( \tau _{e}>t\right) =1/\left( 1+\pi _{2}t\right) $ with
power-law Pareto$\left( 1\right) $ tails.

-c) In the supercritical case $\mu >1$ and $1>\overline{\rho }>0:$

$\mathbf{P}\left( \tau _{e}>t\right) =e^{\left( \mu -1\right) t}/\left(
1+\pi _{2}\left( e^{\left( \mu -1\right) t}-1\right) /\left( \mu -1\right)
\right) \rightarrow \left( \mu -1\right) /\pi _{2}=\overline{\rho }.$
\end{proposition}

\textbf{Proof:} direct resolution. $\Box $

Note $\mathbf{P}\left( \tau _{e}>t\right) =\overline{\rho }+O\left(
e^{-\left( \mu -1\right) t}\right) .$\newline

Consider now \emph{Model} \textbf{2}.
Again we easily establish:

\begin{proposition}
-a) In the subcritical case $\mu <1$and $\overline{\rho }=0:$ we get 
\begin{eqnarray*}
\mathbf{P}\left( \tau _{e}>t\right)  &=&\left[ e^{-\left( \gamma -1\right)
\left( 1-\mu \right) t}/\left( 1+C\left( 1-e^{-\left( \gamma -1\right)
\left( 1-\mu \right) t}\right) /\left( 1-\mu \right) \right) \right]
^{1/\left( \gamma -1\right) } \\
&\sim &\left( 1+C/\left( 1-\mu \right) \right) ^{-1/\left( \gamma -1\right)
}e^{-\left( 1-\mu \right) t},
\end{eqnarray*}
again with exponential tails. The characteristic scale factor is $%
t_{c}=1/\left( 1-\mu \right) $, as in Model $\mathbf{1}$.

-b) In the critical case $\mu =1$ and $\overline{\rho }=0:$

$\mathbf{P}\left( \tau _{e}>t\right) =\left( 1+C\left( \gamma -1\right)
t\right) ^{-1/\left( \gamma -1\right) }$ with power-law Pareto $\left(
1/\left( \gamma -1\right) \right) $ tails. Here the tails of $\tau _{e}$ are
lighter than in Model $\mathbf{1}$, due to $1/\left( \gamma -1\right) >1$.

-c) In the supercritical case $\mu >1$ and $1>\overline{\rho }>0:$

\begin{eqnarray*}
\mathbf{P}\left( \tau _{e}>t\right)  &=&\left[ e^{\left( \gamma -1\right)
\left( \mu -1\right) t}/\left( 1+C\left( e^{\left( \gamma -1\right) \left(
\mu -1\right) t}-1\right) /\left( \mu -1\right) \right) \right] ^{1/\left(
\gamma -1\right) } \\
&\rightarrow &\overline{\rho }=\left( \left( \mu -1\right) /C\right)
^{1/\left( \gamma -1\right) },
\end{eqnarray*}
the probability of explosion$.$
\end{proposition}

\textbf{Proof:} again by direct computation. $\Box $

Note that $\mathbf{P}\left( \tau _{e}>t\right) =\overline{\rho }+O\left(
e^{-\left( \gamma -1\right) \left( \mu -1\right) t}\right) .$ Given $\tau
_{e}<\infty $, the tails of $\tau _{e}$ are exponential with a corrected
scale factor $t_{c}=1/\left[ \left( \mu -1\right) \left( \gamma -1\right)
\right] .$ This fact is in contrast with what was observed in Model $\mathbf{%
1}$.

\subsection{Extreme events and the width of the BGW tree}

In this Section, we shall deal with extreme events pertaining to BGW trees.
To the best of our knowledge these issues have not yet been adressed in the litterature.\newline

Let us first briefly discuss the problem of the largest family size: Let 
\begin{equation*}
M_{t}^{*}=\max \left( M_{1},...,M_{N_{t}}\right)
\end{equation*}
be the maximal offspring number that the $N_{t}$ individuals alive at time $%
t $ can ever give birth to.

Let $F\left( m\right) =\mathbf{P}\left( M\leq m\right) $ be the probability
distribution function of $M$, with $\mathbf{P}\left( M>m\right) =\left[
z^{m}\right] \frac{1-f\left( z\right) }{1-z}$. We have 
\begin{equation*}
\mathbf{P}\left( M_{t}^{*}\leq m\right) =\sum_{n\geq 0}\mathbf{P}\left(
N_{t}=n\right) F\left( m\right) ^{n}=\phi _{t}\left( F\left( m\right)
\right) .
\end{equation*}
Thus, setting $\phi _{t}^{*}\left( m\right) :=\phi _{t}\left( F\left(
m\right) \right) $, $\phi _{t}^{*}\left( m\right) $ is the solution to 
\begin{equation*}
\overset{.}{\phi }_{t}^{*}\left( m\right) =g\left( \phi _{t}^{*}\left(
m\right) \right) ,\text{ }\phi _{0}^{*}\left( m\right) =F\left( m\right) ,
\end{equation*}
which is of the type (\ref{E2}), except for its initial condition.

For the two models under study, this equation can therefore easily be solved, but we
leave the details to the reader.\newline

We now consider another extreme event problem of interest: the maximal value (width of the BGW tree)
that $N_{t}$ can take in its lifetime.
We need to expand the context of our study as follows: so far we have considered a single starting Eve particle.
We now suppose there are $i$ initial particles, each branching independently of the others
according to the same branching mechanism $f$. Then 
\begin{equation*}
\phi _{t}\left( z\right) ^{i}=\mathbf{E}\left( z^{N_{t}}\mid N_{0}=i\right)
\end{equation*}
is the pgf of the whole population size $N_{t}$ at time $t$, given $N_{0}=i$.%
\newline

Let us indeed define $p_{i}\left( k,t\right) =\mathbf{P}\left( N_{s}\leq k\text{ for
all }s\leq t\mid N_{0}=i\right) $ as the probability that, starting from $i$
initial particles, $1\leq i\leq k$, the population size profile keeps
bounded above by $k,$ up to time $t$.

For all $i=1,...,k$, from the Markov property \footnote{%
The following results constitute the continuous-time version of similar
results derived for discrete-time BGW processes in \cite{Adke}.} we
establish time evolution as:

\begin{proposition}
For all $i=1,...,k$, with $p_{0}\left( k,t\right) =1$, we have 
\begin{eqnarray*}
\overset{.}{p}_{1}\left( k,t\right) =-p_{1}\left( k,t\right)
+\sum_{j=1}^{k}\pi _{j}p_{j}\left( k,t\right) +\pi _{0},\text{ }p_{1}\left(
k,0\right) =1, \\
\overset{.}{p}_{i}\left( k,t\right) =-ip_{i}\left( k,t\right)
+i\sum_{j=i-1}^{k}\pi _{j-i+1}p_{j}\left( k,t\right) ,\text{ }p_{i}\left(
k,0\right) =1\text{, }i=2,...,k.
\end{eqnarray*}
\end{proposition}

\textbf{Proof: }We have:

\begin{eqnarray*}
p_{i}\left( k,t\right) &=&e^{-it}+i\int_{0}^{t}dse^{-is}\left(
\sum_{j=0}^{k-i+1}\pi _{j}p_{i+j-1}\left( k,t-s\right) \right) \\
&=&e^{-it}+i\int_{0}^{t}dse^{-is}\sum_{j=i-1}^{k}\pi _{j-i+1}p_{j}\left(
k,t-s\right) \\
&=&e^{-it}\left( 1+i\int_{0}^{t}d\tau e^{i\tau }\sum_{j=i-1}^{k}\pi
_{j-i+1}p_{j}\left( k,\tau \right) \right) .
\end{eqnarray*}
The $e^{-it}$ term arises when the first branching event is larger than $t$
(the pdf of the minimum of $i$ iid exponential$\left( 1\right) $ random
variables), in which case $p_{i}\left( k,t\right) =1$. The second term
arises when the first branching event occurs at $s\leq t,$ in which case,
if the branching particle gives birth to $j$ particles, provided $i+j-1\leq k$%
, $p_{i}\left( k,t\right) $ is given from $p_{i+j-1}\left( k,t-s\right) $
because the new starting number of particles is now $i+j-1$. Note that, if $%
i=1$, this equation exhibits the source term $p_{0}\left( k,\tau \right) =1$
occuring when $j=0$ with probability $\pi _{0}.$

We now rewrite this time evolution in vector form.
Introduce 
\begin{equation*}
\mathbf{p}\left( k,t\right) :=\left( p_{i}\left( k,t\right)
,i=1,...,k\right) ^{\prime }, 
\end{equation*}
with $Q_{k}$ the $k\times k$ upper
Toeplitz-Hessenberg matrix with non-zero entries 
\begin{equation*}
Q_{k}\left( 1,1\right) =1-\pi _{1},\text{ }Q_{k}\left( 1,j\right) =-\pi _{j}%
\text{, }j=2,...,k\text{ and}
\end{equation*}
\begin{eqnarray*}
Q_{k}\left( i,i-1\right) &=&-i\pi _{0},\text{ }Q_{k}\left( i,i\right)
=i\left( 1-\pi _{1}\right) \text{,} \\
Q_{k}\left( i,j\right) &=&-i\pi _{j-i+1}\text{, }i=2,...,k;\text{ }%
j=i+1,...,k,
\end{eqnarray*}
with $\mathbf{r}_{k}^{\prime }=\left( \pi _{0},0,...,0\right) $ \footnote{%
Here and throughout all the paper, a bold $\mathbf{x}$ represents a column
vector with appropriate dimension so that its transpose, say $\mathbf{x}%
^{\prime }$, is a row vector.}, we have the compact algebraic form ($\mathbf{%
1}^{\prime }=\left( 1,...,1\right) $ denotes the unit row vector) 
\begin{equation*}
\overset{.}{\mathbf{p}}\left( k,t\right) =-Q_{k}\mathbf{p}\left( k,t\right) +%
\mathbf{r}_{k},\text{ }\mathbf{p}\left( k,0\right) =\mathbf{1}.
\end{equation*}
Note that, with $\mathbf{k}^{\prime }:=\left( 1,2,...,k\right) $ and $D_{%
\mathbf{k}}=$diag$\left( \mathbf{k}\right) $, $Q_{k}=D_{\mathbf{k}}\overline{%
Q}_{k}$ for some $\overline{Q}_{k}$ involving only the $\pi _{j-i+1}$s and $%
\overline{Q}_{k}=I-P_{k}$ for some substochastic matrix $P_{k}$ with $P_{k}%
\mathbf{1}<\mathbf{1}$. From this structure of $Q_{k}$, this matrix is
invertible with $Q_{k}^{-1}=\left( I-P_{k}\right) ^{-1}D_{\mathbf{k}}^{-1}$
and $\left( I-P_{k}\right) ^{-1}$ a potential matrix. $\Box $

\begin{proposition}
The probability that, starting from $i$ initial particles, the overall
maximum population size keeps bounded above by $k$ for ever, is given by: $%
p_{i}\left( k\right) :=\mathbf{e}_{i}^{^{\prime }}Q_{k}^{-1}\mathbf{r}%
_{k}=\pi _{0}Q_{k}^{-1}\left( i,1\right) $.
\end{proposition}

\textbf{Remark:} In particular, $p_{k}\left( k\right) =\pi
_{0}Q_{k}^{-1}\left( k,1\right) $ is the probability that, starting from $k$
initial particles, the overall maximum population size keeps equal to $k$
for ever.\newline

\textbf{Proof: }The solution of the latter differential equation is

\begin{equation*}
\mathbf{p}\left( k,t\right) =e^{-Q_{\left( k\right) }t}\mathbf{1}%
+\int_{0}^{t}dse_{k}^{-Q_{k}s}\mathbf{r}_{k}=e^{-Q_{k}t}\mathbf{1}+\left(
I-e^{-Q_{k}t}\right) Q_{k}^{-1}\mathbf{r}_{k}.
\end{equation*}
As $t\rightarrow \infty $, $p_{i}\left( k,t\right) \rightarrow \mathbf{e}%
_{i}^{^{\prime }}Q_{k}^{-1}\mathbf{r}_{k}=\mathbf{P}\left( N_{t}\leq k\text{
for all }t\geq 0\mid N_{0}=i\right) $ where $\mathbf{e}_{i}^{^{\prime
}}=(0,...,0,1,0,...,0)$ is the $i$th unit row vector of size $k$, with $1$
in position $i$. Note that $p_{i}\left( k\right) -p_{i}\left( k-1\right) $
is the probability that, starting from $i$ initial particles, the overall
maximum population size is exactly equal to $k.$ $\Box $

We now establish an interesting result on the joint probability of maximum size
and its time-of-reach:

\begin{proposition}
Starting from $1\leq i<k$ particles, the joint probability that the maximum
population size is $k$ and that this maximum value is reached exactly at
time $t$ for the first time is: 
\begin{equation}
\left( p_{i}\left( k-1,t\right) \sum_{j=1}^{k-1}je^{-tQ_{k-1}}\left(
i,j\right) \pi _{k-j+1}\right) p_{k}\left( k\right) .  \label{E4}
\end{equation}
\end{proposition}

\textbf{Proof:}

Let $1\leq i<k$. The term $e^{-tQ_{k-1}}\left( i,j\right) $ is the
probability, starting from $i$ particles, that $N_{t}=j<k$ given $N_{s}\leq
k-1$ for all $s\leq t.$ The term $p_{i}\left( k-1,t\right) $ is the
probability that $N_{s}\leq k-1$ for all $s\leq t$, so the product of the
two is the probability, starting from $i$ particles, that $N_{t}=j<k$ and $%
N_{s}\leq k-1$ for all $s\leq t.$ Recall 
\begin{equation*}
p_{i}\left( k-1,t\right) =\mathbf{e}_{i}^{^{\prime }}\left( e^{-Q_{k-1}t}%
\mathbf{1}+\left( I-e^{-Q_{k-1}t}\right) Q_{k-1}^{-1}\mathbf{r}_{k-1}\right)
.
\end{equation*}

Now, if $t$ is a branching time for any of the $j$ particles alive at $t_{-}$%
, 
\begin{equation*}
p_{i}\left( k-1,t\right) \sum_{j=1}^{k}je^{-tQ_{k-1}}\left( i,j\right) \pi
_{k-j+1}
\end{equation*}
is the probability (density) that the first hitting time of $k$ is $t$ and
that $k$ is the maximal value over the past. Multiplying this probability by 
$p_{k}\left( k\right) $, the probability that, starting from $k$ initial
particles, the overall maximum population size stays lower or equal to $k$ for ever
in the future, and making use of the independence of the past and the future
gives the result. $\Box $\newline

\textbf{Remarks:} $\left( i\right) $ Integrating (\ref{E4}) with respect to $%
t>0$, we obtain that 
\begin{equation*}
p_{i}\left( k\right) -p_{i}\left( k-1\right)
\end{equation*}
is the marginal probability that the maximum population size is $k\geq i$
given $N_{0}=i.$

$\left( ii\right) $ Summing (\ref{E4}) with respect to $k \geq i$ gives the marginal
density of the first hitting time of the maximum over the lifetime.

A second result on size and (this time) overshot time is:

\begin{proposition}
Starting from $1\leq i<k$\ particles, the joint probability that the maximum
population size over the past is $k$\ and that this maximum value is
overshot exactly at time $t$\ for the first time is: 
\begin{equation}
-\overset{.}{p}_{i}\left( k,t\right) p_{k}\left( k\right) =\mathbf{e}%
_{i}^{^{\prime }}e^{-Q_{k}t}\left( Q_{k}\mathbf{1-r}_{k}\right) .  \label{E3}
\end{equation}
\end{proposition}

\textbf{Proof:}

Let $1\leq i<k$. Because with $\tau _{k}=\inf \left( s>0:N_{s}\geq k\mid
N_{0}=i\right) ,$\ defining the first overshooting time of $k$, $(N_{s}\leq k$\ for
all $s\leq t\mid N_{0}=i)$\ $\Rightarrow $\ $\left( \tau _{k}>t\mid
N_{0}=i\right) $, $p_{i}\left( k,t\right) $\ is also $\mathbf{P}\left(
N_{s}\leq k\text{ for all }s\leq t\text{ and }\tau _{k}>t\mid N_{0}=i\right) 
$. We now have 
\begin{equation*}
\overset{.}{p}_{i}\left( k,t\right) =\mathbf{e}_{i}^{^{\prime
}}e^{-Q_{k}t}\left( \mathbf{r}_{k}-Q_{k}\mathbf{1}\right)
\end{equation*}
with 
\begin{eqnarray*}
-\overset{.}{p}_{i}\left( k,t\right) &=&\mathbf{P}\left( N_{s}<k\text{ for
all }s<\tau _{k}\text{ and }\tau _{k}=t\mid N_{0}=i\right) \\
&=&\mathbf{e}_{i}^{^{\prime }}e^{-Q_{k}t}\left( Q_{k}\mathbf{1-r}_{k}\right)
,
\end{eqnarray*}
the joint probability that, given $N_{0}=i<k$, the maximum value of $N_{t}$\
over the past is $k$\ and that the first overshooting time density of this
value $k$\ occurs at $\tau _{k}=t.$\ Note that $Q_{k}\mathbf{1-r}_{k}>%
\mathbf{0}$\ as required from the substochasticity of $P_{k}$, if $-\overset{%
.}{p}_{i}\left( k,t\right) $\ is to be the probability density of some
event. $\Box $\newline

We can now explicitely compute the $Q_{k}^{-1}\left( i,1\right) $, required
for instance in 
\begin{equation*}
\mathbf{P}\left( N_{t}\leq k\text{ for all }t\geq 0\mid N_{0}=i\right) =\pi
_{0}Q_{k}^{-1}\left( i,1\right) ,
\end{equation*}
i.e. the probability that, starting from $i$ initial particles, the overall
maximum population size remains bounded above by $k$ for ever$.$ \newline

This is achieved by introducing the generating function for $\theta$
coefficients as a power series:

\begin{equation*}
\theta \left( z\right) :=\sum_{k\geq 1}\theta _{k}z^{k},
\end{equation*}
with $\theta _{0}=0,$ $\theta _{1}=\left( \left( 1-\pi _{1}\right) /\pi
_{0}\right) ,$ $\theta _{k}=-\left( \pi _{k}/\pi _{0}\right) $, $k\geq 2.$
Of course we first have the trivial determinantal identity 
\begin{equation*}
\left| Q_{k}\right| =k!\left| I-P_{k}\right| .
\end{equation*}
From $\left( 2\right) $ in \cite{Ins}, we thus have the key expansion
property expressing the generating function for the determinants as inverse
of the original generating function: 
\begin{equation*}
\frac{1}{1-\theta \left( z\right) }=\sum_{k\geq 0}\frac{\left| Q_{k}\right| 
}{k!}\left( z/\pi _{0}\right) ^{k}=\sum_{k\geq 0}\left| I-P_{k}\right|
\left( z/\pi _{0}\right) ^{k}.
\end{equation*}
Thus $\left| I-P_{k}\right| =\pi _{0}^{k}\left[ z^{k}\right] \left( \frac{1}{%
1-\theta \left( z\right) }\right) .$ By Faa di Bruno formula (see \cite
{comtet}, p. $137$) one then has: 
\begin{equation*}
\left| I-P_{k}\right| =\pi _{0}^{k}B_{k}\left( \theta _{\bullet }\right)
\end{equation*}
where 
\begin{equation*}
B_{k}\left( \theta _{\bullet }\right) =\sum_{l=1}^{k}B_{k,l}\left( \theta
_{\bullet }\right)
\end{equation*}
are the complete Bell numbers of the sequence $\theta _{\bullet }=\left(
\theta _{1},\theta _{2},...\right) ,$ (see \cite{comtet}, p. $133$),
obtained by summing the ordinary Bell polynomials $B_{k,l}\left( \theta
_{\bullet }\right) $ in the indeterminates $\theta _{\bullet }.$ Now 
\begin{equation*}
Q_{k}^{-1}\left( i,1\right) =\left( I-P_{k}\right) ^{-1}\left( i,1\right) =%
\frac{\left( -1\right) ^{i+1}}{\left| I-P_{k}\right| }C_{1,i},
\end{equation*}
where $C_{1,i}$ is the $\left( 1,i\right) -$cofactor of $I-P_{k}.$ Clearly
now $C_{1,i}=\left( -\pi _{0}\right) ^{i-1}\left| I-P_{k-i}\right| $. Using
this, we finally obtain 
\begin{equation*}
p_{i}\left( k\right) =\pi _{0}Q_{k}^{-1}\left( i,1\right) =\pi _{0}\frac{%
\left( -1\right) ^{i+1}}{\left| I-P_{k}\right| }C_{1,i}=\pi _{0}\left(
-1\right) ^{i+1}\left( -\pi _{0}\right) ^{i-1}\frac{\left| I-P_{k-i}\right| 
}{\left| I-P_{k}\right| }=\frac{B_{k-i}\left( \theta _{\bullet }\right) }{%
B_{k}\left( \theta _{\bullet }\right) },
\end{equation*}
in terms of a ratio of Bell numbers.

Corresponding expressions can be obtained for Model $\mathbf{2}$, while
plugging in the $\pi _{k}$s, as given in (\ref{E1}).\newline

For simplicity let us finally compute the exact values and asymptotic
behaviour of $p_i(k)$ for Model $\mathbf{1}$.

\textbf{In the subcritical case} ($\pi _{0}>\pi _{2}$), with $z_{-}=\pi
_{0}/\pi _{2}$, we have 
\begin{equation*}
\frac{1}{1-\theta \left( z\right) }=\frac{1}{\left( 1-z\right) \left(
1-z/z_{-}\right) }=\frac{A}{1-z}+\frac{B}{1-z/z_{-}}
\end{equation*}
where $A=-z_{-}/\left( 1-z_{-}\right) $ and $B=1/\left( 1-z_{-}\right) $.
Thus 
\begin{equation*}
\left| I-P_{k}\right| =\pi _{0}^{k}\frac{\pi _{0}}{\pi _{0}-\pi _{2}}\left(
1-\left( \frac{\pi _{2}}{\pi _{0}}\right) ^{k+1}\right) ,
\end{equation*}
leading to 
\begin{equation*}
p_{i}\left( k\right) =\pi _{0}Q_{k}^{-1}\left( i,1\right) =\pi _{0}\left(
-1\right) ^{i+1}\left( -\pi _{0}\right) ^{i-1}\frac{\left| I-P_{k-i}\right| 
}{\left| I-P_{k}\right| }=\frac{1-\left( \frac{\pi _{2}}{\pi _{0}}\right)
^{k-i+1}}{1-\left( \frac{\pi _{2}}{\pi _{0}}\right) ^{k+1}}.
\end{equation*}
When $k$ gets large, $p_{i}\left( k\right) \sim 1-\left( \frac{\pi _{2}}{\pi
_{0}}\right) ^{k-i+1}\rightarrow 1$ and $1-p_{i}\left( k\right) $ decreases
geometrically with $k.$ The term $1-p_{i}\left( k\right) $ is the
probability that, starting from $i$ initial particles, the overall maximum
population size overshoots $k$ at least once in the BGW process lifetime
(before $\tau _{e}$).

Let us now consider  \textbf{the critical case} ($\pi _{0}=\pi _{2}$). Setting $%
\pi _{2}=\pi _{0}+\varepsilon $ in the latter formula, we get 
\begin{equation*}
p_{i}\left( k\right) \underset{\varepsilon \rightarrow 0}{\sim }\frac{%
1-\left( 1-\frac{\varepsilon }{\pi _{2}}\left( k-i+1\right) \right) }{%
1-\left( 1-\frac{\varepsilon }{\pi _{2}}\left( k+1\right) \right) }\underset{%
\varepsilon \rightarrow 0}{\sim }1-\frac{i}{k+1}.
\end{equation*}
We conclude that $1-p_{i}\left( k\right) $ decreases algebraically like $i/k$
with $k$, hence much slower than in the subcritical case$.$

\textbf{In the supercritical case} ($\pi _{0}<\pi _{2}$), if the process
explodes, $p_{i}\left( k\right) =0$ and conditioned on non-explosion, we are
taken back to the previous subcritical study with the new branching
mechanism $f_{\rho }\left( z\right) =\rho ^{-1}f\left( \rho z\right) $ where 
$\rho <1$ is the extinction probability solving $f\left( \rho \right) =\rho $%
, here $\rho =\pi _{0}/\pi _{2}$. Thus $f_{\rho }\left( z\right) =\rho
^{-1}f\left( \rho z\right) =\pi _{2}+\pi _{1}z+\pi _{0}z^{2}$, exchanging
the roles of $\pi _{0}$ and $\pi _{2}.$

\section{Spatial branching process in dimension 1}

We have until now dealt with zero-space dimension tree-like branching processes. Let us move to
aspects of the spatial BGW
process, first of all restricted to one-dimensional case.  We shall revisit some results of \cite{Sawyer} and extend
them to a new situation akin to Model $2$. In such a spatial branching
process, an Eve particle diffuses according to one-dimensional standard
Brownian motion (with diffusion constant fixed to $1$ without loss of
generality). At some (mean one) exponential time, it dies out giving birth
in the process to $M$ offspring; if $M>0$, the daughter particles diffuse
according to independent standard Brownian motions, started where the mother
particle died.

Let $p\left( x\right) $ be the probability that the Eve particle starting at 
$0$ has some descendant ever diffusing above the threshold $x\geq 0.$ Then 
\cite{Sawyer} $p\left( x\right) $ solves ($p\left( 0\right) =1$) 
\begin{equation}
\frac{1}{2}p^{\prime \prime }-h\left( p\right) =0\text{ or }p^{\prime
}\left( x\right) ^{2}-4\int^{p\left( x\right) }h\left( z\right) dz=Cte,
\label{eq1}
\end{equation}
as a stationary solution of the Kolmogorov-Petrovsky-Piskounov equation, 
\cite{KPP}.

We note that $p\left( x\right) $ is also the probability that the supremum
of the positions of all particles that appeared at any time exceeds $x$, so $%
p\left( x\right) =1-q\left( x\right) $ where $q\left( x\right) $ is a
probability distribution function (which in particular is monotone
non-decreasing).

Because $p\left( x\right) \underset{x\rightarrow \infty }{\rightarrow }%
\overline{\rho }$, the limit $p^{\prime }\left( x\right) $ should also exist
and this limit is necessarily $0.$ These equations are then generically solved by inverting the quadrature:

\begin{equation*}
x=\frac{1}{2}\int_{p\left( x\right) }^{1}H\left( y\right) ^{-1/2}dy,
\end{equation*}
where $H\left( y\right) =\int_{\overline{\rho }}^{y}h\left( z\right) dz.$
Let us investigate $p\left( x\right) $ for the previous two examples:\newline

Consider first \emph{Model} \textbf{1}.\newline

\begin{proposition}
-a) In the subcritical case $\mu <1$ and $\overline{\rho }=0:$

$H\left( y\right) =\frac{\left( 1-\mu \right) }{2}y^{2}\left( 1+\frac{%
2\pi _{2}y}{3\left( 1-\mu \right) }\right) $. Then
\begin{equation*}
p\left( x\right) =\frac{4AE\left( x\right) }{c\left( 1-AE\left( x\right)
\right) ^{2}}
\end{equation*}
where $E\left( x\right) =\exp \left( -\sqrt{2\left( 1-\mu \right) }x\right) ,
$ $c=\frac{2\pi _{2}}{3\left( 1-\mu \right) }$ and $A=\left( \sqrt{1+c}%
-1\right) ^{2}/c.$

Note that $p\left( x\right) $ has exponential tails with scale factor $%
x_{c}=1/\sqrt{2\left( 1-\mu \right) }$ and $p\left( 0\right) =1$.\newline

-b) In the critical case $\mu =1$ and $\overline{\rho }=0:$ Here $p\left(
x\right) =\left( 1+x/x_{c}\right) ^{-2}$ hence $p\left( x\right) $ decays
algebraically at infinity with exponent $2.$ The scale factor is $x_{c}=%
\sqrt{3/\pi _{2}}/4$.\newline

-c) In the supercritical case $\mu >1$ and $1>\overline{\rho }>0:$ Here,
with $\overline{\rho }=1-\pi _{0}/\pi _{2},$ $p\left( x\right) =\overline{%
\rho }+O\left( e^{-x/x_{c}}\right) ,$ where $x_{c}=1/\sqrt{2h^{\prime
}\left( \overline{\rho }\right) }$.
\end{proposition}

\textbf{Proof:}

-a) and -b) follow by direct computations. Concerning -c), $p\left( x\right) 
$ has an atom at infinity which is the probability of explosion of the
underlying branching process and the remaining tails are exponential.
Indeed, letting $p\left( x\right) -\overline{\rho }=\widetilde{p}\left(
x\right) $, $\widetilde{p}\left( x\right) $ solves $\frac{1}{2}\widetilde{p}%
^{\prime \prime }-h\left( \widetilde{p}+\overline{\rho }\right) =0$ which
for small $\widetilde{p}$ (large $x$) is $\frac{1}{2}\widetilde{p}^{\prime
\prime }-h^{\prime }\left( \overline{\rho }\right) \widetilde{p}=0$,
recalling $h\left( \overline{\rho }\right) =0.$ One can check that $%
h^{\prime }\left( \overline{\rho }\right) =\pi _{2}-\pi _{0}>0$ and so $%
\widetilde{p}$ is exponential with the right scale factor$.$ $\Box $\newline

As observed in \cite{Sawyer}, any branching model for which $f\left(
z\right) =\pi _{0}+\pi _{1}z+\pi _{2}z^{2}+O\left( z^{\gamma ^{\prime
}}\right) $ with $\gamma ^{\prime }>2,$ will display similar tail behaviors.%
\newline

Let us now move to \emph{Model} \textbf{2}.We get:\newline

\begin{proposition}
-a) In the subcritical case $\mu <1$and $\overline{\rho }=0:$

$H\left( y\right) =\frac{\left( 1-\mu \right) }{2}y^{2}\left( 1+\frac{%
2Cy^{\gamma -1}}{\left( \gamma +1\right) \left( 1-\mu \right) }\right) $. One computes:
\begin{equation*}
p\left( x\right) =\left( \frac{4AE\left( x\right) }{c\left( 1-AE\left(
x\right) \right) ^{2}}\right) ^{1/\left( \gamma -1\right) }
\end{equation*}
where:

$E\left( x\right) =\exp \left( -\left( \gamma -1\right) \sqrt{2\left( 1-\mu
\right) }x\right) ,$ $c=\frac{2C}{\left( \gamma +1\right) \left( 1-\mu
\right) }$ and $A=\left( \sqrt{1+c}-1\right) ^{2}/c;$ $p\left( x\right) $
has again exponential tails with scale factor $x_{c}=1/\sqrt{2\left( 1-\mu
\right) }$.\newline

-b) In the critical case $\mu =1$ and $\overline{\rho }=0:$ We get $p\left(
x\right) =\left( 1+x/x_{c}\right) ^{-2/\left( \gamma -1\right) }$ and $%
p\left( x\right) $ decays algebraically at infinity with exponent $2/\left(
\gamma -1\right) .$ The scale factor is $x_{c}=\sqrt{\left( \gamma +1\right)
/C}/\left( \gamma -1\right) $.\newline

-c) In the supercritical case $\mu >1$ and $1>\overline{\rho }>0:$ Setting $%
\overline{\rho }=\left( \left( \mu -1\right) /C\right) ^{1/\left( \gamma
-1\right) },$ one has $p\left( x\right) =\overline{\rho }+O\left(
e^{-x/x_{c}}\right) ,$ where $x_{c}=1/\sqrt{2h^{\prime }\left( \overline{%
\rho }\right) }$. $p\left( x\right) $ has an atom at infinity which is the
probability of explosion (non-extinction) and the remaining tails are
exponential.
\end{proposition}

\textbf{Proof:}

Statements -a) and -b) are obtained by direct computations. To get statement -c),
setting as before $p\left( x\right) -\overline{\rho }=\widetilde{p}\left(
x\right) $, $\widetilde{p}\left( x\right) $ solves $\frac{1}{2}\widetilde{p}%
^{\prime \prime }-h\left( \widetilde{p}+\overline{\rho }\right) =0$ which
for small $\widetilde{p}$ (large $x$) is $\frac{1}{2}\widetilde{p}^{\prime
\prime }-h^{\prime }\left( \overline{\rho }\right) \widetilde{p}=0$,
recalling $h\left( \overline{\rho }\right) =0.$ One checks that here $%
h^{\prime }\left( \overline{\rho }\right) =\left( \mu -1\right) \left(
\gamma -1\right) >0$ and $\widetilde{p}$ is exponential with the claimed
scale factor, different from the scale factor obtained in the subcritical
case$.$ $\Box $\newline

The quantity $p\left( x\right) $ is equivalently the probability that an Eve
particle started at $x\geq 0$ has some descendant ever diffusing below the
threshold $x=0.$ This way of thinking $p\left( x\right) $ also pertains to
dimensions not equal to one which we move to now.

\section{Spatial branching process in dimension $d\neq 1$}

\label{section2}Let $P\left( \mathbf{x}\right) $ be the probability that
some particle started at $\mathbf{x}$ in $\Bbb{R}^{d}$ ($d=2,3,...$) has
some descendant ever diffusing within a ball of radius $\varepsilon >0$
around the origin, with $x:=\left\| \mathbf{x}\right\| _{2}=\left(
\sum_{i=1}^{d}x_{i}^{2}\right) ^{1/2}>\varepsilon .$ Then, from \cite{Sawyer}%
, introducing $\Delta $ as the $d-$dimensional Laplacian, $P\left( \mathbf{x}%
\right) $ solves $\frac{1}{2}\Delta P-h\left( P\right) =0$ and in view of
rotational invariance, $p\left( x\right) :=P\left( \left\| \mathbf{x}%
\right\| _{2}\right) $ solves 
\begin{equation}
\frac{1}{2}p^{\prime \prime }+\frac{d-1}{2x}p^{\prime }-h\left( p\right) =0,
\label{Equ1}
\end{equation}
We impose the boundary conditions $p\left( \varepsilon \right) =1$ and $%
p\left( \infty \right) =0$ for consistency with the probabilistic
interpretation of $p$. Indeed this modified construction is dictated by the
fact that $d-$dimensional branching Brownian motion with $d=2,3,...$ has
zero probability to meet the origin. $\frac{1}{2}\partial _{x}^{2}+\frac{d-1%
}{2x}\partial _{x}$ is the Bessel generator of the modulus of a $d-$%
dimensional Brownian motion. Thus $p\left( x\right) $ is the probability
that the full trail of the $d-$dimensional branching Brownian motion ever
happened to be at distance to the origin less than $\varepsilon $. Hence the
boundary conditions.

This construction can be extended to non-integer $d (\geq 2)$ as follows. Let $%
R_{t}=\exp \left( B_{t}+at\right) $ and $X_{t}=R_{\tau _{t}}$ where $\tau
_{t}=\int_{0}^{t}X_{s}^{-2}ds$ and $B_{t}$ is the standard Brownian motion.
Assume $a\geq 0$. Then the infinitesimal generator of $X_{t}>0$, as a
time-changed geometric Brownian motion $R$ with non-negative drift, is \cite
{GP} 
\begin{equation*}
\frac{1}{2}\partial _{x}^{2}+\frac{\left( 2a+1\right) }{2x}\partial _{x},
\end{equation*}
so it is the generator of some Bessel process (say BS$_{d}$), with
`dimension' parameter $d=2\left( a+1\right) \geq 2$, not necessarily an
integer.

\begin{proposition}
Denote by $p\left( x\right) $ the probability that some branching $1-$%
dimensional BS$_{d}$ particle system, started at $x>0,$ has some descendant
ever diffusing below $\varepsilon $ ($x>\varepsilon >0$). Then $p\left(
x\right) $ solves 
\begin{equation*}
\frac{1}{2}p^{\prime \prime }+\frac{d-1}{2x}p^{\prime }-h\left( p\right) =0,
\end{equation*}
with $p\left( \varepsilon \right) =1$ and $p\left( \infty \right) =0.$
\end{proposition}

The BS$_{d}$ process $X$ is well-defined even if $d>1$ then with $\mathbf{E}%
\left( \int_{0}^{t}ds/X_{s}\right) <\infty $, and also even if $d>0,$ \cite
{Yor}$.$ We also recall some basic properties of BS$_{d}$ processes with
respect to their dimension $d$ as from \cite{RY}:

- For $d>2$, the process BS$_{d}$ is transient.

- For $d\geq 2$ the point $0$ is polar and for $d\leq 1$ it is reached
almost surely.

- For $0<d<2$, BS$_{d}$ is recurrent (null recurrent if $d\in \left(
1,2\right] $, positive recurrent otherwise); the point $0$ is
instantaneously reflecting.

We are now in a position to extend the interpretation of the latter
differential equation describing the BS$_{d}$ process $X$ when $d$ is non
integer. Integer values of $d$ all correspond to a $d-$dimensional Brownian
motion with full rotational invariance and the occurrence of $d$ in the
differential equation follows from the reduction of a $d-$dimensional
Laplacian to invariant configurations. We consistently conjecture that
non-integer values of $d$ similarly characterize Brownian motion on a
fractal-type background (possibly relevant in epidemics propagation
description) again with a full `rotational'' invariance, here by the
simplest analytic continuation of the differential equation to non-integer
values of $d$ (More complicated analytic extensions involving additional,
real-periodic functions may be considered but they shall not be addressed
here). This bears some technical resemblance with procedures in quantum
field theories such as dimensional regularization. This conjecture is
strongly borne out by the following checks: starting from the definition of
spherically symmetric random walks in non-integer dimensions $d$ by Bender
et al. \cite{BBM} and taking the large (continuous) limit of radii of the
nested $d$-dimensional spheres between which the particle random-walks, one
recovers exactly the drift contribution $\frac{d-1}{2x}$ and the constant
unit local variance term in the second-order differential operator
generating $BS_{d}$. Following this interpretation, the $BS_{d}$ process $%
X_{t}$ may be viewed as the modulus of some isotropic $d-$dimensional
diffusion process, evolving in a $d-$dimensional space for which the surface
of a ball with radius $x$ is $2\pi ^{d/2}x^{d-1}/\Gamma \left( d/2\right) $. 
\newline

We shall now study equation (\ref{Equ1}) when $h\left( z\right) =\left(
1-\mu \right) z+Cz^{\gamma }$, $\gamma \in \left( 1,2\right) $. It is not
solvable contrary to the $d=1$ case, except for \textit{integer} values of $%
\gamma$ (elliptic functions for $\gamma = 3$, hyperelliptic functions for $%
\gamma = 4, 5 \cdots$. These integer values however lie beyond the interval
of relevance for the probabilistic interpretation of the model. It must be
however suggested that rational values of $\gamma$ lying in the relevant
open interval $\left( 1,2\right)$ may still lead to solutions with some
interpretation in algebraic geometry (multiple coverings of elliptic or Prym
manifolds).

Anyway here we limit ourselves to the sole asymptotic analysis (large $x$)
form of the solutions with the suitable limit behavior $p\left( \infty
\right) =0$ (except in one case).

The case $\gamma =2$ (Model \textbf{1}) has been analyzed to a large extent
(in the critical regime) by \cite{Sawyer}. Some further extensions of the
binary branching model (either subcritical or critical) has also been
reported in \cite{DMRZ}, in the $d=2$ dimensional case, involving a deep
study of the dynamics of both perimeter and area of the convex hull of the
BBM trail.

\subsection{Subcritical case ($\mu <1$)}

Recalling again the asymptotic limit behavior $p\left( \infty \right) =0$ we
conclude that the higher power term $Cz^{\gamma }$ is to be dropped when
analyzing around $\infty $. The large $x$ (small $p$) solutions are thus
governed by 
\begin{equation*}
\frac{1}{2}p^{\prime \prime }+\frac{d-1}{2x}p^{\prime }-\left( 1-\mu \right)
p=0,
\end{equation*}
which can be mapped into a modified Bessel equation, (see \cite{Bow}, p. $117
$). Indeed, with $\alpha ,\beta $ some constants, let 
\begin{equation*}
p\left( x\right) =x^{\alpha }J_{\alpha }\left( \beta x\right) ,
\end{equation*}
where $J_{\alpha }\left( x\right) $ obeys $J_{\alpha }^{\prime \prime
}+J_{\alpha }^{\prime }/x+\left( 1-\alpha ^{2}/x^{2}\right) J_{\alpha }=0$,
as a Bessel function of the first kind, of order $\alpha .$ Then $p$ obeys 
\begin{equation*}
p^{\prime \prime }-\frac{2\alpha -1}{x}p^{\prime }+\beta ^{2}p=0.
\end{equation*}
Setting $\alpha =\left( 2-d\right) /2$ and $\beta =i\sqrt{2\left( 1-\mu
\right) }=:i\gamma $ and recalling that $I_{\alpha }\left( x\right)
=i^{-\alpha }J_{\alpha }\left( ix\right) $ is the modified Bessel function
of the first kind of order $\alpha $, we get:

\begin{equation*}
p\left( x\right) =x^{\alpha }\left( A_{1}I_{\alpha }\left( \gamma x\right)
+A_{2}K_{\alpha }\left( \gamma x\right) \right) ,
\end{equation*}
where $K_{\alpha }\left( x\right) $is the modified Bessel function of the
second kind of order $\alpha $.

Recalling $K_{\alpha }\left( x\right) \sim e^{-x}\sqrt{\pi /\left( 2x\right) 
}$ near $x=\infty $ and keeping only the decaying factor at $\infty $, we
establish:

\begin{proposition}
In the subcritical case, $p(x)$ behaves for large $x$ as 
\begin{equation*}
p\left( x\right) \sim A_{2}x^{\alpha }K_{\alpha }\left( \gamma x\right) \sim
\lambda x^{-\left( d-1\right) /2}e^{-\sqrt{2\left( 1-\mu \right) }x}\text{, }%
\lambda >0
\end{equation*}
\end{proposition}

Compared to the exact $d=1$ case studied before, the asymptotics of $p\left(
x\right) $ exhibit an extra $x^{-\left( d-1\right) /2}$power term.

\subsection{Critical case ($\mu =1$)}

In this case, $h\left( p\right) =Cp^{\gamma }$.
A conformal covariance property then arises:

\begin{proposition}
If $p$ is a solution of (\ref{Equ1}) with $h\left( p\right) =Cp^{\gamma }$,
then, for all $\lambda >0$, $p_{\lambda }\left( x\right) =\lambda ^{2/\left(
\gamma -1\right) }p\left( \lambda x\right) $ are also solutions. The
constant $2/\left( \gamma -1\right) $ is the conformal weight for $p$.
\end{proposition}

This will play an important role in the next analysis. In particular the conformal invariant
solution $m\left( x\right) =x^{-2/\left( \gamma
-1\right) }, $ will appear.

\subsubsection{Behavior of $p$ near infinity}

We first assume an asymptotic power-law form $p\sim \lambda x^{-\alpha }$, $%
\alpha ,\lambda >0$, leading to 
\begin{equation*}
\lambda \alpha x^{-\left( \alpha +2\right) }\left( \alpha +2-d\right)
-2C\lambda ^{\gamma }x^{-\gamma \alpha }=0.
\end{equation*}

Let us first analyze the power-law behaviour. 
We need to impose $\alpha +2\leq \gamma \alpha ,$ otherwise the dominant term would
be the unique one $x^{-\gamma \alpha }$, which would be inconsistent. So $%
\alpha \geq 2/\left( \gamma -1\right) .$\newline

- Suppose first $\alpha >2/\left( \gamma -1\right) .$ Then necessarily the
power-law exponent is $\alpha =d-2$ and this regime occurs when $%
d>d_{c}:=2+2/\left( \gamma -1\right) .$ Note that there is no specification
of what $\lambda $ is (except of course for $\lambda >0$).\newline

- Suppose now $\alpha =2/\left( \gamma -1\right) .$ Then the two power terms
contribute equally likely and the solution asymptotically behaves like the
conformally invariant monomial $m\left( x\right) =x^{-2/\left( \gamma
-1\right) },$ obeying $m\left( x\right) =\lambda ^{2/\left( \gamma -1\right)
}m\left( \lambda x\right) $.\newline

We can now discuss the scale factor $\lambda$ .\newline

* Suppose first $d\neq \alpha +2$; then we also need to have $%
d<d_{c}=2+\alpha =2+2/\left( \gamma -1\right) $ in addition with 
\begin{equation*}
\lambda \alpha +\left( \alpha +2-d\right) =2C\lambda ^{\gamma },
\end{equation*}
leading to $\lambda =\left( \alpha \left( \alpha +2-d\right) /C\right)
^{1/\left( \gamma -1\right) }=\left( \left( d_{c}-d\right) /\left( \left(
\gamma -1\right) C\right) \right) ^{1/\left( \gamma -1\right) }.$\newline

* Suppose now $d=\alpha +2$; then $d=d_{c}$ and we have to try the enhanced
asymptotic form $p\sim \lambda x^{-\alpha }\left( \log x\right) ^{\beta }$, $%
\alpha =d_{c}-2$, $\lambda >0.$ We get 
\begin{eqnarray*}
\frac{d-1}{x}p^{\prime } &\sim &\lambda \left( d_{c}-1\right) x^{-\left(
\alpha +2\right) }\left( \beta \left( \log x\right) ^{\beta -1}-\alpha
\left( \log x\right) ^{\beta }\right) \\
p^{\prime \prime } &\sim &\lambda x^{-\left( \alpha +2\right) }\left( \alpha
\left( \alpha +1\right) \left( \log x\right) ^{\beta }-\beta \left( 2\alpha
+1\right) \left( \log x\right) ^{\beta -1}+\beta \left( \beta -1\right)
\left( \log x\right) ^{\beta -2}\right) .
\end{eqnarray*}
Plugging these estimates into $p^{\prime \prime }+\frac{d-1}{x}p^{\prime
}-2h\left( p\right) =0,$ the $\left( \log x\right) ^{\beta }$ terms cancel,
leading to 
\begin{equation*}
-\alpha \lambda \beta x^{-\left( \alpha +2\right) }\left( \log x\right)
^{\beta -1}-2C\lambda ^{\gamma }\left( \log x\right) ^{\beta \gamma
}x^{-\alpha \gamma }=0,
\end{equation*}
discarding the $\left( \log x\right) ^{\beta -2}$ term as compared to $%
\left( \log x\right) ^{\beta -1}$. Observing $\alpha \gamma =\alpha +2,$
this can be achieved only if $\beta \gamma =\beta -1,$ so if $\beta
=-1/\left( \gamma -1\right) .$ The constant $\lambda $ is also determined by 
$\lambda ^{\gamma -1}=-\alpha \beta /\left( 2C\right) $, so $\lambda =\left(
C\left( \gamma -1\right) ^{2}\right) ^{-1/\left( \gamma -1\right) }$.\newline

To summarize:

\begin{proposition}
In the critical case, the behavior of $p(x)$ for large $x$ depends on the
value of the dimension $d$ with respect to a critical dimension $d_{c}:=2+2/\left(
\gamma -1\right) $.

\begin{itemize}
\item  if $d>d_{c}$, $p\sim \lambda x^{-\left( d-2\right) }$ with $\lambda >0
$ being left unspecified.

\item  if $d=d_{c}$, $p\sim \lambda \left( x^{-2}/\log \left( x\right)
\right) ^{\left( d_{c}-2\right) /2}$ with $\lambda =\left( C\left( \gamma
-1\right) ^{2}\right) ^{-1/\left( \gamma -1\right) }$.

\item  if $d<d_{c}$, $p\sim \lambda x^{-2/\left( \gamma -1\right) }$ with $%
\lambda =\left( \left( d_{c}-d\right) /\left( \left( \gamma -1\right)
C\right) \right) ^{1/\left( \gamma -1\right) }$.
\end{itemize}
\end{proposition}

\subsubsection{Behavior of $p$ near the origin}

Although this question does not necessarily make sense in our probabilistic
context because $p\left( x\right) $ is intrinsically defined for $x>\varepsilon $%
, it turns out that the formal analysis of $p\left( x\right) $ near the
origin is possible.

We first try the asymptotic power-law form $p\sim \lambda x^{-\beta }$, $%
\beta ,\lambda >0$, leading to 
\begin{equation*}
\lambda \beta x^{-\left( \beta +2\right) }\left( \beta +2-d\right)
-2C\lambda ^{\gamma }x^{-\gamma \beta }=0.
\end{equation*}

Once again we first analyze the power-law behaviour.
We need to impose $\beta +2\geq \gamma \beta ,$ otherwise the dominant term would
be the unique one $x^{-\gamma \beta }$, which would fail. So $\beta \leq
2/\left( \gamma -1\right) .$\newline

- Suppose first $\beta <2/\left( \gamma -1\right) .$ Then necessarily the
power-law exponent is $\beta =d-2.$ This regime occurs when $%
2<d<d_{c}:=2+2/\left( \gamma -1\right) .$ Note that there is again no
specification of what $\lambda $ is (except of course for $\lambda >0$).%
\newline

- Suppose now $\beta =2/\left( \gamma -1\right) .$ Then the two power terms
contribute equally likely and we are back to the conformally invariant
solution.\newline

Let us now discuss the scale factor $\lambda$.\newline

* Suppose first $d\neq \beta +2$; then we also need to have $d<d_{c}=2+\beta
=2+2/\left( \gamma -1\right) $ in addition with 
\begin{equation*}
\lambda \beta +\left( \beta +2-d\right) =2C\lambda ^{\gamma },
\end{equation*}
leading to $\lambda =\left( \beta \left( \beta +2-d\right) /C\right)
^{1/\left( \gamma -1\right) }=\left( \left( d_{c}-d\right) /\left( \left(
\gamma -1\right) C\right) \right) ^{1/\left( \gamma -1\right) }.$\newline

* Suppose now $d=\beta +2$; then $d=d_{c}$ and we try the asymptotic form $%
p\sim \lambda x^{-\beta }\left( -\log x\right) ^{\delta }$, $\beta =d_{c}-2$%
, $\lambda >0.$ We get 
\begin{eqnarray*}
\frac{d-1}{x}p^{\prime } &\sim &\lambda \left( d_{c}-1\right) x^{-\left(
\beta +2\right) }\left( -\delta \left( -\log x\right) ^{\delta -1}-\beta
\left( -\log x\right) ^{\delta }\right) \\
p^{\prime \prime } &\sim &\lambda x^{-\left( \beta +2\right) }\left( \beta
\left( \beta +1\right) \left( -\log x\right) ^{\delta }+\delta \left( 2\beta
+1\right) \left( -\log x\right) ^{\delta -1}+\delta \left( \delta -1\right)
\left( -\log x\right) ^{\delta -2}\right) .
\end{eqnarray*}
Plugging these estimates into $p^{\prime \prime }+\frac{d-1}{x}p^{\prime
}-2h\left( p\right) =0,$ the $\left( -\log x\right) ^{\beta }$ terms cancel
again, leading to 
\begin{equation*}
\lambda \beta \delta x^{-\left( \beta +2\right) }\left( -\log x\right)
^{\delta -1}-2C\lambda ^{\gamma }\left( -\log x\right) ^{\delta \gamma
}x^{-\beta \gamma }=0,
\end{equation*}
discarding the $\left( -\log x\right) ^{\delta -2}$ term compared to $\left(
-\log x\right) ^{\delta -1}$ when $x$ is small. Observing $\beta \gamma
=\beta +2,$ this could be achieved only if $\delta \gamma =\delta -1,$ so if 
$\delta =-1/\left( \gamma -1\right) .$ The constant $\lambda $ should also
be determined by $\lambda ^{\gamma -1}=\beta \gamma /\left( 2C\right) $ and
because $\beta <0$, $\lambda $ cannot be real \footnote{%
There exist solutions with a complex prefactor which we disregard, due to
their lack of physical meaning so far, in particular because it hampers an
interpretation of $p$ as a probability.}.\newline

Interestingly enough we may also define consistent solutions of the alternative 
asymptotic form $p\sim c\left( 1+\lambda
x^{\beta }\right) $ for some constants $c,\beta >0.$ To leading order, we
need to have 
\begin{equation*}
\lambda c\beta \left( \beta +d-2\right) x^{\beta -2}=2c^{\gamma }
\end{equation*}
which also requires $\beta =2$ together with $\lambda =c^{\gamma -1}/d>0.$%
\newline

To summarize:

\begin{proposition}
In the critical case, the behavior of $p(x)$ near $x=0$ also depends on the
value of the dimension $d$ with respect to the critical dimension $d_{c}$ (with $%
d_{c}=2+2/\left( \gamma -1\right) $)

\begin{itemize}
\item  if $2<d<d_{c}$, $p\sim \lambda x^{-\left( d-2\right) }$ with $\lambda
>0$ being left unspecified.

\item  if $0<d<d_{c}$, $p\sim \lambda x^{-2/\left( \gamma -1\right) }$ with $%
\lambda =\left( \left( d_{c}-d\right) /\left( \left( \gamma -1\right)
C\right) \right) ^{1/\left( \gamma -1\right) }$.

\item  if $d>0$, $p\sim c\left( 1+\lambda x^{2}\right) $ with $c>0$ and $%
\lambda =c^{\gamma -1}/d$.

\item  if $d=d_{c}$, there is no real solution of the form $\lambda
x^{-\beta }\left( -\log x\right) ^{\delta }$ with $\beta ,\delta ,\lambda $
real. A solution nevertheless exists, albeit with $\lambda$ complex.

\item  if $d>d_{c}$, there is no real solution either.
\end{itemize}
\end{proposition}

\subsection{Supercritical case ($\mu >1$)}

We must slightly modify the asymptotic behavior at infinity in this case by substracting
a non-zero asymptotic limit corresponding to the zero of the potential term.
Defing accordingly $\overline{\rho }=\left( \left( \mu -1\right) /C\right) ^{1/\left(
\gamma -1\right) },$ let $p\left( x\right) -\overline{\rho }=\widetilde{p}%
\left( x\right) $. Then $\widetilde{p}\left( x\right) $ solves $\frac{1}{2}%
\widetilde{p}^{\prime \prime }+\frac{d-1}{2x}-h\left( \widetilde{p}+%
\overline{\rho }\right) =0$ which for small $\widetilde{p}$ (large $x$) is $%
\frac{1}{2}\widetilde{p}^{\prime \prime }+\frac{d-1}{2x}-h^{\prime }\left( 
\overline{\rho }\right) \widetilde{p}=0$, recalling $h\left( \overline{\rho }%
\right) =0.$ Recall $h^{\prime }\left( \overline{\rho }\right) =\left( \mu
-1\right) \left( \gamma -1\right) >0.$

The large $x$ (small $\widetilde{p}$) solutions of $\widetilde{p}$ are thus
governed by 
\begin{equation*}
\frac{1}{2}\widetilde{p}^{\prime \prime }+\frac{d-1}{2x}\widetilde{p}%
^{\prime }-h^{\prime }\left( \overline{\rho }\right) \widetilde{p}=0,
\end{equation*}
which can be mapped into a modified Bessel equation as before, but now with $%
\beta =i\gamma $ and $\gamma =\sqrt{2h^{\prime }\left( \overline{\rho }%
\right) }$.

Proceeding similarly as in the subcritical case, we now get ($\lambda >0$)

\begin{equation*}
\widetilde{p}\left( x\right) \sim A_{2}x^{\alpha }K_{\alpha }\left( \gamma
x\right) \sim \lambda x^{-\left( d-1\right) /2}e^{-\sqrt{2\left( \mu
-1\right) \left( \gamma -1\right) }x}.
\end{equation*}
Finally, we obtained

\begin{proposition}
In the supercritical case, $p(x)$ behaves for large $x$ as 
\begin{equation*}
p\left( x\right) \sim \left( \left( \mu -1\right) /C\right) ^{1/\left(
\gamma -1\right) }+\lambda x^{-\left( d-1\right) /2}e^{-\sqrt{2\left( \mu
-1\right) \left( \gamma -1\right) }x}.
\end{equation*}
\end{proposition}

Again, as compared to the $d=1$ case studied before, the asymptotics of $%
p\left( x\right) $ has an extra $x^{-\left( d-1\right) /2}$ power factor in
the corrective term $\widetilde{p}\left( x\right) $.\newline

Let us supply a final result pertaining to the supercritical regime:
conditionally given the extinction time is finite, the underlying branching
process is subcritical with offspring pgf $f_{\rho }\left( z\right) :=\rho
^{-1}f\left( \rho z\right) ,$ obeying $f_{\rho }\left( 1\right) =1$, $%
f_{\rho }^{\prime }\left( 1\right) =f^{\prime }\left( \rho \right) <1.$

Recalling $f\left( z\right) =1-\mu \left( 1-z\right) +C\left( 1-z\right)
^{\gamma }$ where $\mu /\gamma >C>\mu -1$ and $\gamma \in \left( 1,2\right)
, $ we indeed get $f_{\rho }^{\prime }\left( 1\right) =f^{\prime }\left(
\rho \right) =\mu -C\gamma \overline{\rho }^{\gamma -1}=\mu -\gamma \left(
\mu -1\right) <1.$ Defining $h_{\rho }\left( z\right) :=f_{\rho }\left(
1-z\right) -\left( 1-z\right) $, we get 
\begin{equation*}
h_{\rho }\left( z\right) =\left( 1-\mu \right) z+C\frac{\overline{\rho }%
^{\gamma }}{\rho }\left( \left( 1+\frac{\rho }{\overline{\rho }}z\right)
^{\gamma }-1\right) ,
\end{equation*}
which is regular near $z=1$.

We have $f_{\rho }\left( z\right) \sim _{z=0}\rho ^{-1}\left( 1-\mu
+C\right) +\left( \mu -C\gamma \right) z+\rho C\gamma \left( \gamma
-1\right) z^{2}/2+O\left( z^{3}\right) ,$ so we are in the domain of
attraction of the subcritical model studied in Section $3.1$. Defining $\mu
_{\rho }:=\mu -\gamma \left( \mu -1\right) $, and applying the results of
Section $3.1$, we conclude that

\begin{proposition}
Conditionally given that the supercritical branching process survives 
\begin{equation*}
p\left( x\right) \sim \frac{\lambda }{\overline{\rho }}x^{-\left( d-1\right)
/2}e^{-\sqrt{2\left( 1-\mu _{\rho }\right) }x}=\frac{\lambda }{\overline{%
\rho }}x^{-\left( d-1\right) /2}e^{-\sqrt{2\left( \gamma -1\right) \left(
\mu -1\right) }x}\text{, }\lambda >0,
\end{equation*}
displaying the modified scale factor $x_{c}=1/\sqrt{2\left( \gamma -1\right)
\left( \mu -1\right) }.$
\end{proposition}

\textbf{Remark:} Let: 
\begin{equation*}
x_{V}=\left( \frac{\pi ^{d/2}}{\Gamma \left( d/2+1\right) }\right) ^{-1/d}%
\text{ and }x_{S}=\left( \frac{2\pi ^{d/2}}{\Gamma \left( d/2\right) }%
\right) ^{-1/\left( d-1\right) }.
\end{equation*}
If the epidemics starts at distance $x$ of the origin, the tail probability
of its spatial extension in $d-$dimensional space for which the volume of a
ball is $V=\left( x/x_{V}\right) ^{d}$ will be: $\mathbf{P}(V>v)\sim
p(x_{V}v^{1/d}),$ $v>0$ large, where the large $x$ behaviors of $p(x)$ are
given in Propositions $10-12$ and $14$. Similarly, the tail probability of
the area of the boundary $S$ of the $d-$dimensional sphere $V$ will be $%
\mathbf{P}(S>s)\sim p(x_{S}v^{1/\left( d-1\right) }),$ $s>0$ large.

\textbf{Acknowledgments:} T. H. is indebted to Satya Majumdar (LPTMS, Orsay)
for bringing to his attention the paper \cite{Sawyer}. T.H. also
acknowledges partial support from the labex MME-DII (Mod\`{e}les
Math\'{e}matiques et \'{E}conomiques de la Dynamique, de l' Incertitude et
des Interactions).


\begin{thebibliography}{99}
\bibitem{Adke}  Adke, S. R. The maximum population size in the first $N$
generations of a branching process. Biometrics 20, (1964), 649-651.

\bibitem{Bow}  Bowman, F. Introduction to Bessel functions. Dover
Publications Inc., New York, 1958.

\bibitem{BBM} Bender C.M. ; Boettcher S. and Moshe M. Spherically-Symmetric
Random Walks in Noninteger Dimension. J. Math. Phys. 35, (1994), 4941-4963.

\bibitem{comtet}  Comtet, L. Advanced Combinatorics. D.~Reidel Publishing
Company, Dord\-recht, Holland, 1974.

\bibitem{DMRZ}  Dumonteil, E.; Majumdar, S. N.; Rosso, A. and Zoia, A.
Spatial extent of an outbreak in animal epidemics. PNAS (2013), vol. 110 no.
11, 4239-4244.

\bibitem{Yor}  G\"{o}ing-Jaeschke, A.; Yor, M. A survey and some
generalizations of Bessel processes. Bernoulli 9 (2003), no. 2, 313-349.

\bibitem{GP}  Graversen, S. E.; Peskir, G. Maximal inequalities for Bessel
processes. J. Inequal. Appl. 2 (1998), no. 2, 99-119.

\bibitem{Harris}  Harris, T. E. The theory of branching processes. Die
Grundlehren der Mathematischen Wissenschaften, Bd. 119 Springer-Verlag,
Berlin; Prentice-Hall, Inc., Englewood Cliffs, N.J. 1963.

\bibitem{Ins}  Inselberg, A. On determinants of Toeplitz-Hessenberg matrices
arising in power series. J. Math. Anal. Appl. 63 (1978), no. 2, 347-353.

\bibitem{KPP}  Kolmogorov, A.; Petrovsky, I. and Piskounov, N. Etude de l\'{}%
\'{e}quation de la diffusion avec croissance de la quantit\'{e} de mati\`{e}%
re et son application \`{a} un probl\`{e}me biologique. Moscou Univ. Bull.
Math. 1, pp. 125, (1937).

\bibitem{Lamperti}  Lamperti, J. An occupation time theorem for a class of
stochastic processes. Trans. Amer. Math. Soc. 88, (1958), 380-387.

\bibitem{RY}  Revuz , D.; Yor, M. Continuous martingales and Brownian motion,
volume 293 of Grundlehren der Mathematischen Wissenschaften [Fundamental
Principles of Mathematical Sciences]. Springer-Verlag, Berlin,1991.

\bibitem{Sawyer}  Sawyer, S.; Fleischman J. Maximum geographic range of a
mutant allele considered as a subtype of a Brownian branching random field,
PNAS, USA, Vol. 76, no 2, (1979), pp. 872-875.
\end{thebibliography}
\end{document}